# 의미적으로 중요한 시각적 내용의 온토로지 기반의 보안검색

Ontology-based Secure Retrieval of Semantically Significant Visual Contents

칸 무함마드, 이르판 메흐무르, 이미영, 지수미, 백성욱[1]

Khan Muhammad, Irfan Mehmood, Mi Young Lee, Su Mi Ji, Sung Wook Baik

(143-747) 서울특별시 광진구 능동로 209 세종대학교 디지털콘텐츠 연구소
Khanmuhammad@sju.ac.kr, irfanmehmood@sju.ac.kr, miylee@sejong.ac.kr,
smji@sejong.ac.kr, sbaik@sejong.ac.kr

## 요 약

이미지 분류는 관심이 높은 연구 분야로 이미지 데이터의 방대한 양을 시각적 콘텐츠에 따라 다양한 클래스로 분류한다. 연구자들은 다른 카테고리에 이미지를 분류하기 위한 다양한 하위 수준 특징 기반 기술을 제시하고 있다. 그러나, 효율적이고 효과적인 분류 및 검색은 여전히 시각적인 내용의 복잡한 특성으로 인해 어려운 문제로 남아있다. 또한, 기존의 정보검색기술은 보안에 취약하여, 환자의 기록 및 법 집행 기관의 데이터베이스와 같은 개인의 시각적 콘텐츠를 제3자에 의해 검색이 쉬웠다. 그러므로, 우리는 보안 이미지 분류 및 정보 검색에 대한 이미지 스테가노그래피를 사용하여 새로운 온톨로지 기반의 프레임 워크를 제안한다. 제안된 프레임워크는 하위 계층 이미지 특징을 효율적인 분류 결과인 온톨로지의 상위 계층 컨셉의 매핑을 위해 도메인 특징 온톨로지를 사용한다. 또한 제안된 방법은 내용 안에 비밀 메시지와 같은 이미지의 의미를 숨기고 정보 검색 프로세스를 제3자로부터 보호하기 위해 이미지 스테가노그래피를 이용한다. 제안된 프레임워크는 기존기술의 복잡도를 최소화하고 개인의 이미지 데이터베이스로부터 안전하고 실시간 시각적 콘텐츠 검색을 위해 적합한 기술을 증대시킨다. 다른 최신 기술의 시스템과 비교하여 실험한 결과 제안된 프레임 워크의 효율성, 유효성 및 보안을 확인 하였다.

## Abstract

Image classification is an enthusiastic research field where large amount of image data is classified into various classes based on their visual contents. Researchers have presented various low-level features-based techniques for classifying images into different categories. However, efficient and effective classification and retrieval is still a challenging problem due to complex nature of visual contents. In addition, the traditional information retrieval techniques are vulnerable to security risks, making it easy for attackers to retrieve personal visual contents such as patient's records and law enforcement agencies' databases. Therefore, we propose a novel ontology-based framework using image steganography for secure image classification and information retrieval. The proposed framework uses domain-specific ontology for mapping the low-level image features to high-level concepts of ontologies which consequently results in efficient classification. Furthermore, the proposed method utilizes image steganography for hiding the image's semantics as a secret message inside them, making the information retrieval process secure from third parties. The proposed framework minimizes the computational complexity of traditional techniques, increasing its suitability for secure and real-time visual contents retrieval from personalized image databases. Experimental results confirm the efficiency, effectiveness, and security of the proposed framework as compared with other state-of-the-art systems.

---

1) Corresponding author

키워드: 정보검색, 온톨로지, 이미지 스테가노그라피, 정보 보안
Keyword: Information retrieval, ontology, image steganography, information security

## 1. Introduction

Image analysis and retrieval is one of the hot research areas of information retrieval. Researchers have presented various image analysis techniques and information retrieval systems[1, 2], belonging to two main categories. The first category is keywords/text meta-data based methods[3, 4], where the visual contents are searched based on user input query, consisting of textual description. This type of searching method have been effectively used by the world famous search engines including Bing and Google[5]. Keywords-based retrieval methods can provide better results in some situations but their accuracy is not consistent due to various reasons such as incorrect spelling during image description process, lack of knowledge for describing a given image in a natural language, difficulty in finding suitable keywords for effective image description, and ignorance of image's features, resulting in redundant and irrelevant visual contents retrieval[2]. The second category of information retrieval is contents-based image retrieval (CBIR) where the images are retrieved based on their features from large databases, resolving the limitations of traditional text-based methods[2]. Researchers from the last few decades have used various low-level image features for effective image retrieval such as color, texture, shape, and spatial location. The well-known color features that have been used for retrieval purposes, include color-moments, color-coherence vector, color-histogram, and color-covariance matrix that are calculated using various human perception-oriented color spaces such as red-green-blue (RGB), YCbCr, hue-saturation-value (HSV), lightness with a-b as color components (LAB), and LUV[5]. The texture features include directionality, regularity, contrast, coarseness, line-likeness, and roughness that can be effectively used in image classification for describing real-world visual contents including trees, fruit skin, fabric, clouds, and bricks and are not well-defined like color features[6]. The shape features include aspect ratio, circularity, Fourier descriptors, and moment's invariants and can be used in various applications, requiring man-made objects[7]. The spatial location features are useful for region classification and show the location of an object in a given image[8].

Although, the low-level features based techniques have been widely used in current CBIR systems, yet they cannot fully describe the high-level semantic concepts of users, leading to semantic gap problem, which consequently reduces the efficiency and effectiveness of current CBIR systems. Furthermore, the traditional techniques don't consider the security aspects of CBIR systems, making it easy for attackers to retrieve personalized records of sensitive databases such as medical repositories and law-enforcement authorities' databases.

To reduce the semantic gap, semantic technologies such as ontology can provide promising results as it can effectively map the low-level image features to high-level concepts of ontology. Therefore, in this paper, we propose an ontology-based framework for secure visual contents retrieval. The major contributions of this work are included, but not limited to:

I. An efficient ontology-based framework for effective image classification and visual contents retrieval is proposed, reducing the computational complexity of traditional CBIR systems.
II. The proposed visual contents retrieval framework is made secure based on image steganography by hiding the image's semantics inside images.
III. A light-weight edges-based data hiding algorithm is proposed for embedding the semantic information inside images, increasing the payload while preserving the visual transparency of stego images
IV. To increase the security, the semantic information is divided into four sub-blocks and are encrypted using an encryption algorithm, making the semantics extraction more challenging for unauthorized users, hence ensuring the security of visual contents retrieval systems.
V. An efficient and light-weight ranking algorithm is proposed for adaptively ranking the retrieved contents into low-level, medium-level, and high-level categories according to users' preferences, making the visualization of retrieved contents easy for further processing and decision making.

The rest of the paper is structured as follows. Section 2 describes the proposed methodology in detail along with its major components including ontology-based visual semantic modeling, edges based steganographic algorithm, semantic extraction algorithm, and ranking algorithm. Section 3 presents the experimental results and discussion. In section 4, the paper is concluded and some future directions are drawn.

## 2. The proposed methodology

In this section, we present a detailed description of the proposed framework for secure visual contents retrieval. First, we present the ontology-based visual semantic modeling. The concept of semantic technology such as ontology plays a vital role in the proposed information retrieval system in mapping the low-level image features to high-level ontology concepts. Then, we discuss a block division procedure and encryption algorithm, increasing the security of embedded information, which consequently makes the information retrieval more secure. Next, we present a data hiding steganographic algorithm, followed by semantic extraction algorithm in detail. Finally, we present the ranking algorithm to rank the desired retrieved images based on user preferences. The major sections of the proposed framework are shown in Fig. 1.

## 2.1 Ontology-based visual sematic modeling

In this section, we present the concept of ontology-based visual semantic modeling in the context of the proposed visual contents retrieval framework. The motivational factor of using sematic technologies (ontology) in the proposed method is its suitability for mapping the low-level features of images into high-level semantics of ontology concepts. This results in accurate classification of large amount of visual contents, which consequently increases the effectiveness and efficiency of the proposed information retrieval framework.

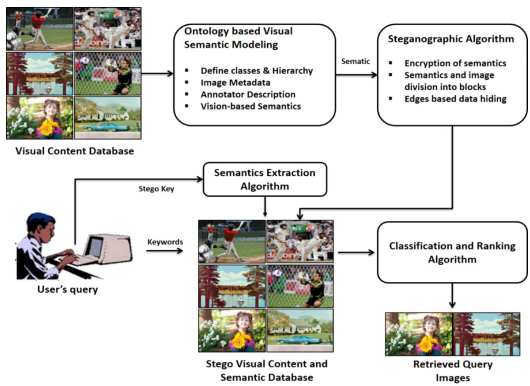

Fig.1 The proposed framework for secure visual contents retrieval

## 2.2 Steganographic Algorithm

In this sub-section, we discuss the detail of the proposed steganographic algorithm for semantic hiding in visual contents. Steganographic algorithms aim to hide secret messages inside innocent-looking carriers such as images, audio, and video such that the existence of hidden data is known to the communicating bodies only and is not detectable using human visual system (HVS)[9]. In the proposed framework, we have used image steganography for hiding the semantics of images inside them, resulting in efficient and secure visual contents retrieval as compared to other low-level features based retrieval systems. The main bedrocks of the proposed steganographic algorithm are shown in Fig. 2 and are described in detail in the next coming sections.

### A. Blocks division and encryption algorithm

This section describes the blocks division of input image and image's semantics, followed by an encryption procedure using an encryption key, increasing the security of embedded semantics. The concept of blocks division and encryption is taken from our recently published work[10], introducing various barriers in the way of attackers. The motivational reason behind its usage is to increase the security of embedded semantics, making the extraction of sensitive semantics more challenging which consequently makes the proposed visual contents retrieval secure from adversaries.

### B. Edges based semantic hiding algorithm (ESHA)

In this section, we present the actual process of hiding the encrypted semantic information into the rotated four sub-blocks. The basic idea of the proposed data hiding method is based on the locality of pixels in the input image i.e. smooth pixels and edgy pixels.

Larger difference between two consecutive pixels show that the pixel is located at edge area and large number of bits can be embedded and vice versa[11]. The motivational fact behind the usage of edges based hiding is to increase the payload i.e. hide more information. In addition, HVS is less sensitive to edge area pixels and hence the information embedded in edge areas cannot be easily detected as compared to smooth areas' pixels. Fig. 3 shows a sample of input image, its corresponding edgy images using various edge detectors, and the final stego image.

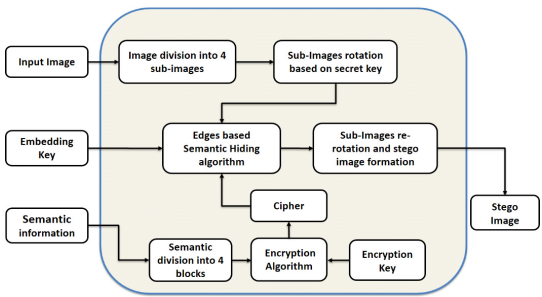

Fig.2 Framework of the proposed semantic hiding algorithm

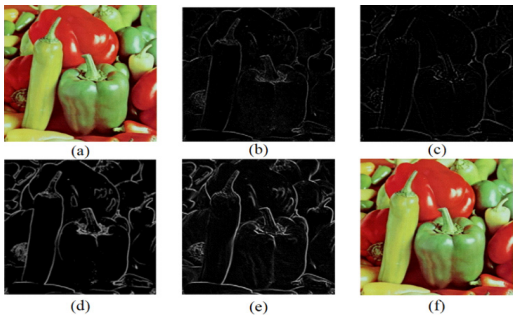

Fig. 3 A sample input image, its various edge-detected versions, and the final stego image. (a) Input image peppers, (b) image with edges detected by Laplacian edge detector, (c) Laplacian of Gaussian (LOG) edge detector result, (d) Fuzzy edge detector result, (e) Hybrid edge detector, resulted by combining (b), (c), and (d) images, and (f) final pepper stego image with embedded semantic information.

## 2.3 Semantics extraction algorithm

In this section, we describe the mechanism of recovering the hidden semantics using the proposed semantic extraction algorithm. When the user makes a query for a given image or class of images along with the keywords and the concerned stego key, the proposed extraction algorithm recovers the embedded semantics and description from the stego image database. The recovered encrypted information is then decrypted using a decryption key and the desired visual contents are retrieved. The semantic extraction process is made secure by introducing various levels of security including stego key for recovery sequence, decryption key for decoding the encrypted contents, and edges based extraction algorithm, maintaining the integrity, secrecy, and confidentiality of visual contents, hence making the visual contents retrieval process secure from security risks.

## 2.4 Ranking algorithm

This sub-section demonstrates the working and importance of the proposed ranking algorithm in our main framework of image classification and retrieval. The ranking algorithm receives a set of images from the semantic extraction algorithm, retrieved from the huge visual contents based on user's query. The retrieved images are classified and ranked into three different categories based on its weightage and user preferences including high-level, medium-level, and low-level ranking. The motivation of using the ranking algorithm is to present the retrieved results adaptively according to the user preferences, hence making the proposed framework more suitable for real-world applications, requiring quick decision making such as military and law enforcement authorities. To make the idea of the proposed ranking algorithm clear, we have presented a simple example. Fig. 4 shows a sample of retrieved images in three categories based on a sample user query as "Sky Images" from the dataset.

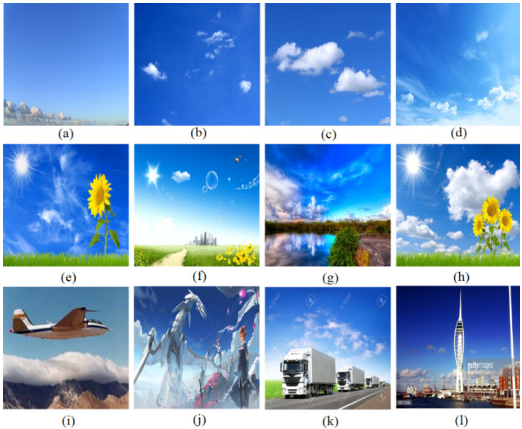

Fig. 4 A sample example of retrieved images classified into three categories including high-level, medium-level, and low-level ranks. The results are based on a user query with the keyword "Sky Images". The first row shows four sample images, belonging to high-level rank. The second row shows a sample of retrieved images, belonging to medium-level rank. The last row is about low-level rank images.

## 3. Experimental Results and Discussion

In this section, we present the experimental setup of the proposed framework and other competing state-of-the-art methods. The proposed method is compared with various competing methods in terms of efficiency, computational complexity, security, and visual quality of stego images. MATLAB version R2013a has been used as a simulation tool for conducting the various experiments. The following sub-sections explain the detail of dataset, performance evaluation metrics, and various experiments conducted.

### 3.1 Dataset

This section demonstrates the detail of the datasets that have been used for performance evaluation. One of the major critical points in evaluation of CBIR systems is selecting a suitable dataset. This is due to the fact that currently no standard dataset for CBIR systems exists and all researchers of this area are not agreed on the number and type of images in a given standard dataset[12]. In this paper, we have used two datasets: COREL database[13] and USC-SIPI-ID[14]. The reason for using the COREL dataset is its wide coverage of semantic groups with a total of 1000 images, having 10 semantic classes including buses, food, horses, beach, flowers, buildings, mountains, dinosaurs, African people in village, and elephants. Each class is further divided into 100 images of dimension 256×384 or 384×256 pixels based on human perception of image similarity. The USC-SIPI-ID dataset is used to measure the visual quality of images in the area of image steganography in particular and image processing in general. A total of 50 standard images have been selected from this dataset with different natures such as edgy and smooth images including Lena, airplane, peppers, baboon, trees, home, and couple. A few sample images from both the datasets are given in Fig. 5.

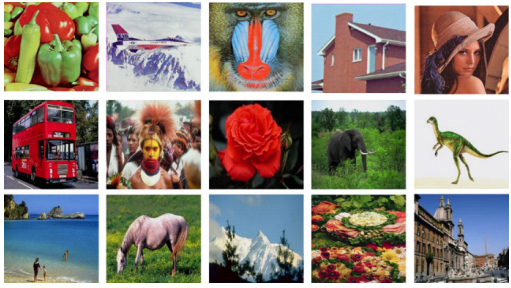

Fig. 5 Sample images from both the datasets. The first row shows the standard images of USC-SIPI-ID dataset including peppers, airplane, baboon, building, and Lena. The second and third row show candidate images from ten semantic groups of the COREL dataset including buses, African people, flower, elephant, dinosaur, beach, horses, mountains, food, and buildings.

### 3.2 Performance evaluation and discussion

In this section, we evaluate the performance of the proposed method and other competing methods using various image quality assessment metrics (IQAMs) and time complexity. The metrics include peak-signal-to-noise-ratio (PSNR) and structural similarity index metric (SSIM) which can be calculated using equation 1-3 as follows [15, 16]:

$$PSNR = 10\log_{10}\left(\frac{C_{max}^2}{MSE}\right) \quad (1)$$

$$MSE = \frac{1}{MN}\sum_{x=1}^{M}\sum_{y=1}^{N}(S_{xy} - C_{xy})^2 \quad (2)$$

$$SSIM(C,S) = \frac{(2\mu_x\mu_y + C_1)(2\sigma_{xy} + C_2)}{(\mu_x^2 + \mu_y^2 + C_1)(\sigma_x^2 + \sigma_y^2 + C_2)} \quad (3)$$

PSNR calculates the amount of distortion in stego images after intentionally hiding the semantic information. It is measured in terms of decibel (dB). The greater the value of PSNR, the better the visual quality is and vice versa. Sometimes, PSNR fails to consider the complete structural information while measuring the similarity, therefore, we have also used the SSIM for evaluation. The value of SSIM lies between 1 and 0. The quantitative results based on PSNR and SSIM for the proposed method and other competing methods including least significant bit (LSB), cyclic steganographic technique (CST) [17], LSB matching (LSB-M) [11], LSB-M revisited (LSB-MR) [18], pixel indicator technique (PIT) [19], Karim's method [20], and HSI-MLSB [21] are given in Fig. 6 and Fig. 7.

Fig. 6 and Fig. 7 show the quantitative experimental results based on two well-known IQAMs including PSNR and SSIM. The results are computed over 50 standard color images and their average score has been used for comparison. It is clear from both the figures that the performance of Karim's method, PIT, and CST is almost same in terms of PSNR and SSIM. The average score of PSNR for LSB, LSB-M, LSB-MR, and HSI-MLSB is relatively same but based on SSIM, there is obvious variation in their average scores. The proposed method dominates all the mentioned state-of-the-art methods by achieving the highest PSNR and SSIM score, validating its effectiveness and better performance.

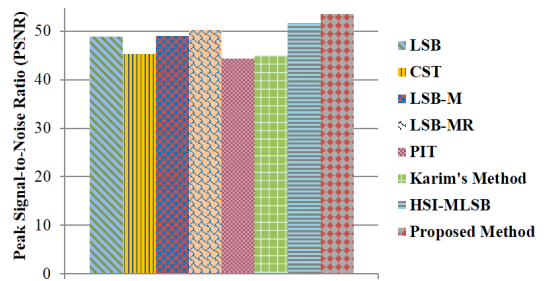

Fig. 6 Qualitative evaluation based on PSNR over 50 standard images

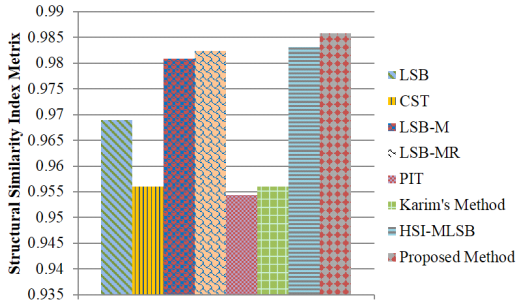

Fig. 7 Qualitative evaluation based on SSIM over 50 standard images

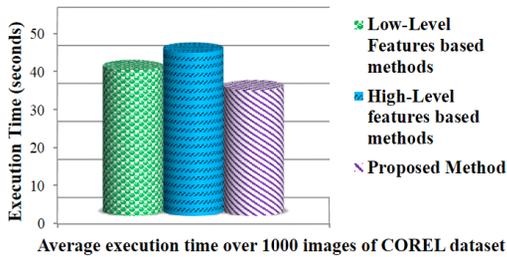

Fig. 8 Efficiency measurement in terms of execution time

Fig. 8 demonstrates the comparison of the proposed framework with low-level and high-level features based visual contents retrieval methods based on execution time in seconds. The comparison is based on 1000 images of COREL dataset. From Fig. 8, it can be confirmed that the proposed method reduces the computational complexity of low-level and high-level features based methods, hence making it more suitable for real-time applications such as surveillance systems and law enforcement authorities' systems.

## 4. Conclusion and Future Work

In this paper, an efficient ontology-based secure visual contents retrieval model is proposed. The proposed framework uses domain-specific ontologies to map the low-level image's features to high-level ontology concepts, making the classification and visual contents retrieval more efficient. The proposed retrieval framework is made secure using the concept of image steganography by embedding the image's semantics inside them. To increase the security, the semantics and input image are divided into four blocks and each message block is embedded into one of the four image blocks using a secret pattern. Furthermore, the semantics are encrypted prior to data hiding, making the extraction of embedded semantics more challenging for attackers without the required authentic information. The proposed framework reduces the computational complexity of traditional retrieval systems, making it more suitable for secure and real-time visual contents retrieval in medical and law-enforcement authorities' databases. In future work, the authors tend to integrate the concept of image encryption with image steganography for increasing the security of top-sensitive visual contents. The efficiency will be further improved by embedding a feature vector of various semantic information instead of textual description.

## Acknowledgments

This research is supported by the ICT R&D program of MSIP/IITP. [2014(R0112-14-1014), the Development of Open Platform for Service of Convergence Contents].

## Author

◆ **Khan Muhammad**

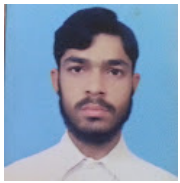

- He received his BS degree in Computer Science from Islamia College Peshawar, Pakistan. He is currently pursuing Master course in Sejong University, Seoul, Korea. His research interests include digital image and video processing, Information Hiding and Security.

◆ **Irfan Mehmood**

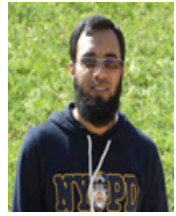

- He received his BS degree in Computer Science from National University of Computer and Emerging Sciences from Pakistan. He is currently pursuing his Ph.D. degree at Sejong University, Seoul, Korea. His major research interests include visual information summarization and prioritization, in which he explored several new directions of semantically significant information extraction in surveillance and medical image / videos.

◆ **Mi Young Lee**

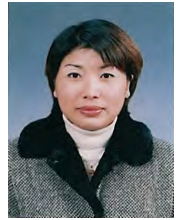

- She is a research professor at Sejong University. She received her PhD degree in the Image and Information Engineering at Pusan National University. Her research interests include Interactive Contents, UI, UX and Developing Digital Contents. She has a MS in the Department of Image and Information Engineering from Pusan National University.

◆ **Su Mi Ji**

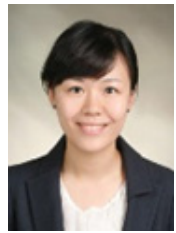

- She is a postdoctoral researcher in the Institute of Digital Contents at Sejong University. She received her PhD degree in Digital Contents from Sejong University, Seoul, South Korea. Her research interests include Contents Authoring tool, Data Visualization, UX and Serious Game. She has a MS in the Department of Educational Game from Kwangwoon University.

◆ **Sung Wook Baik**

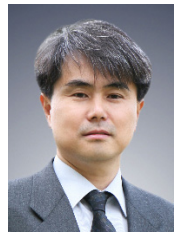

- He is a professor in the College of Electronics and Information Engineering at Sejong University. His research interests include Computer vision, Pattern recognition, Computer game and AI. He is a PhD in Information Technology and Engineering from George Mason University.